\documentstyle[proc]{npbproc}

\newcommand{\nn}{\nonumber}

\newcommand{\nd}{\noindent}

\newcommand{\bq}{\begin{equation}}
\newcommand{\eq}{\end{equation}}
\newcommand{\ba}{\begin{eqnarray}}
\newcommand{\ea}{\end{eqnarray}}
\newcommand{\ra}{\rightarrow}
\newcommand{\MSb}{$\overline{\mbox{MS}}$}
\newcommand{\DISg}{$\mbox{DIS}_{\gamma}$}

\begin{document}

\title{PHOTON STRUCTURE: QCD TREATMENT AND PARTON DENSITIES$\, ^{\ast }$}
\author{Andreas Vogt}
\address{Sektion Physik, Universit\"at M\"unchen, Theresienstra{\ss}e 37,
	 D--80333 Munich, Germany}

\runtitle{Photon structure: QCD treatment and parton densities}
\runauthor{A. Vogt}

\begin{abstract}
The QCD treatment of the photon structure beyond the leading order is discussed
with emphasis on a proper choice of the factorization scheme and/or the
boundary conditions. Recent parametrizations of the photon's parton content
are examined and compared. The sensitivity of the photon structure function
on the QCD scale parameter is reconsidered.
\end{abstract}

\maketitle

\section{Introduction}
The hadronic structure of real photons \cite{BeWa}, see also \cite{Ze94}, is
most easily illustrated by considering inclusive electron-photon
deep-inelastic scattering (DIS). This process is completely analogous to the
usual lepton-nucleon DIS. Consequently, the unpolarized cross section for
$ e\gamma \ra e\gamma^{\star} \gamma \ra eX $ can be expressed in terms of two
structure functions $ F_{1,2}^{\,\gamma}(x,Q^{2}) $, where $x$ denotes the
Bjorken scaling variable and $ Q^{2} $ is the virtuality of the probing
off-shell photon. Only $ F_{2}^{\,\gamma} $ has been measured so far.
The main issues addressed in this talk can already be introduced at the parton
model level. Here, two different contributions occur, which originate in
pointlike and non-pointlike photon-quark couplings.

The pointlike contribution derives from the lowest order QED process
(`Born-Box') $ \gamma^{\star} \gamma \rightarrow q \bar{q} $ \cite{WZ}, and
is for $ Q^{2} \gg m_{q}^{2} $ given by
\bq
  \frac{1}{x}F_{2,B}^{\,\gamma} = \frac{\alpha}{2\pi} \sum_{q} 2 e_{q}^{4}
    \big[ k_{q}^{\, (0)}(x) \ln \frac{Q^{2}}{m_{q}^{2}} + C_{\gamma}(x) \big]
  \:\: .
\eq
Effective masses $ m_{q} $ of the quarks with charges $ e_{q} $ had to be
introduced in order to avoid collinear singularities. Contrary to all hadrons
the $x$-dependence of $ F_{2}^{\,\gamma} $, governed by $ k_{q}^{\, (0)}(x) $
and $ C_{\gamma}(x) $, is calculable and the structure function rises linearly
with $ \ln Q^{2} $.
Analogous to the usual $ \overline{\mbox{MS}} $ treatment of hadronic DIS,
the quark densities in the photon can be defined by absorbing the universal
mass singularity,
\bq
  q_{B,\,\overline{MS}}^{\gamma} = \bar{q}_{B,\,\overline{MS}}^{\,\gamma} =
    \frac{\alpha}{2\pi} e_{q}^{2} \, 3 \big[ x^{2} + (1-x)^{2} \big]
    \ln \frac{Q^{2}}{m_{q}^{2}} \: ,
\eq
where the concrete form of $ k_{q}^{\, (0)} $ has been put in. The $
C_{\gamma } $-term, not absorbed into the parton distributions in (2), then
acts as the subleading, `direct' contribution to $ F_{2}^{\,\gamma } $.
However, this separation creates problems in the next-to-leading order QCD
treatment, mainly since $ C_{\gamma } $ is negative and divergent at $ x \ra
1 $. Solutions to this problem, either by introducing a different factorization
scheme (`\DISg ') or by adding some `technical' \MSb\ input for the evolution
equations, are examined in section 2.

The hadronic, non-pointlike contribution to $ F_{2}^{\,\gamma} $ can be
illustrated by the well-known coupling of the photon to the vector mesons
$ \rho $, $ \omega $ and $ \phi $ (vector meson dominance, VMD) \cite{BeWa},
resulting in
\bq
  F_{2,VMD}^{\,\gamma} = \big(4\pi \alpha / f_{\rho}^{2} \big) F_{2}^{\rho}
    + \ldots \:\: .
\eq
This part is completely analogous to the behaviour of hadrons: there is no rise
with $ \ln Q^{2} $ and the $x$-shape is not calculable in perturbation theory.
Hence unknown input parton distributions enter the description just as in
hadron structure. In section 3 recent parametrizations of $ q^{\gamma} $ and
the less well known gluon density $ G^{\,\gamma} $ are discussed and compared.

The dependence of $ F_{2}^{\,\gamma} $ on the unknown effective quark masses
is removed by the inclusion of the dominant QCD corrections, leading to the
famous large-$x$ large-$ Q^{2} $ parameter-free asymptotic QCD predictions
\cite{Wit,BB}. These results, however, are not appropriate at scales accessible
at present or in the near future, $ Q^{2} \stackrel{<}{{\scriptstyle \sim}} 300
\mbox{ GeV}^{2} $ \cite{GGR}. The original hope on an especially clean $
\alpha_{S} $ determination from $ F_{2}^{\,\gamma} $ has therefore been
dampend. Some debate has arisen on how much sensitivity survives in a proper,
non-asymptotic treatment including the uncalculable hadronic boundary
conditions, see e.g.\ \cite{VS}. A short note is added to this discussion in
section 4.
\section{QCD: Next to Leading Order}
The generalized evolution equations for the $ {\cal O}(\alpha ) $ quark and
gluon content of the photon are \cite{BB,DWi}
\ba
  \frac{dq^{\gamma}}{d\ln Q^{2}} &=& \frac{\alpha}{2\pi} e_{q}^{2} k_{q} +
  \frac{\alpha_{S}} {2\pi}\, \big[ P_{qq} \otimes q^{\gamma} + \ldots \big]
                                                              \nonumber  \\
  \frac{dG^{\,\gamma}}{d\ln Q^{2}} &=& \frac{\alpha}{2\pi}k_{G} +
  \frac{\alpha_{S}} {2\pi}\, \big[ P_{GG} \otimes G^{\,\gamma} + \ldots \big]
  \:\: .
\ea
Here $ \otimes $ denotes the usual convolution, and the normal hadronic part
has not been written out completely. The photon distribution in the photon,
$ \Gamma^{\gamma} = \delta (1-x) - {\cal O}(\alpha ) $, has dropped out at
the order of the electromagnetic coupling $ \alpha $ considered here,
resulting in the characteristic inhomogeneous $k$-terms. A salient consequence
of (4) is the absence of a momentum sum rule interrelating $ q^{\gamma} $ and
$ G^{\, \gamma} $. Hence an important constraint on the gluon density, present
in the hadronic case, is missing here. The splitting functions $ P_{ij} $ and
$ k_{i} $ are up to now known to next-to-leading order (NLO) accuracy in the
strong coupling $ \alpha_{S} \equiv \alpha_{S}(Q^{2}) $, see \cite{FKL} and
\cite{FoPi,GRVt}, respectively.

The singlet distributions can be decomposed as
\bq
  \vec{q}^{\,\,\gamma} = \left( \!\! \begin{array}{c} 2 \sum_{q} q^{\gamma}
    \\ G^{\,\gamma} \end{array} \!\! \right)
    = \vec{q}_{PL}^{\,\,\gamma} + \vec{q}_{had}^{\,\,\gamma} \:\: ,
\eq
where $ q^{\gamma} \! = \! \bar{q}^{\,\gamma} $ has been used. The well-known
homo\-geneous hadronic solution $ \vec{q}_{had}^{\,\,\gamma} $ contains the
perturbatively uncalculable boundary conditions $ \vec{q}^{\,\,\gamma}
(Q_{0}^{2}) $.  The inhomogeneous (`pointlike', PL) solution then satisfies
$ \vec{q}_{PL}^{\,\,\gamma}(Q_{0}^{2}) = 0 $.

An analytical solution of (4) is possible for Mellin-moments. The NLO
pointlike part reads \cite{GR,GRVt}
\ba
  \vec{q}^{\,\,\gamma}_{PL} &=& \Big\{ \frac{1}{\alpha_{S}} + \hat{U}\Big\}
      \Big\{ 1-\big[\alpha_{S}/\alpha_{S}(Q_{0}^{2})\big]^{1+\hat{d}} \Big\}
      \frac{1}{1+\hat{d}} \, \vec{a}
                                                           \nonumber  \\
  & & \mbox{} + \Big\{ 1-\big[\alpha_{S}/\alpha_{S}(Q_{0}^{2})\big]^{\hat{d}}
      \Big\}  \frac{1}{\hat{d}} \, \vec{b} + {\cal O}(\alpha_{S}) \:\: ,
\ea
with $\vec{a} $, $\vec{b} $, $\hat{d} $ and $\hat{U} $ being known Mellin-$n$
dependent combinations of the splitting functions.  The $x$-dependent
distributions are obtained by a numerical Mellin-inversion of (6) and the
corresponding hadronic and non-singlet expressions.
A major advantage of this procedure is that inconsistent higher order
$ {\cal O}(\alpha_{S}) $ terms, which would lead to spurious $ x \ra 1 $
singularities, can be trivially omitted. In a direct NLO $x$-space calculation,
their cancellation requires somewhat cumbersome algorithms \cite{Rossi,Drees}.
\begin{figure}[tb]
\vspace{9.9cm}
\caption{The pointlike structure function $ F_{2,\, PL}^{\,\gamma} $ in LO and
in NLO for the \MSb\ and \DISg\ factorization schemes. Also shown is the
modified $ F_{2,\, PL^{\prime}}^{\,\gamma} $ of \protect\cite{AFG} (AFG).
$ Q_{0}^{2} = 1 \mbox{ GeV}^{2}$, three active flavours and $ \Lambda = 0.2
\mbox{ GeV} $ have been used.}
\end{figure}

The structure function $ F_{2}^{\, \gamma} $ is at NLO given by
\ba
  \frac{1}{x} F_{2}^{\,\gamma} & = & \sum 2e_{q}^{2} \Big\{ q^{\gamma}
      +  \frac{\alpha_{S}}{2\pi} (C_{q} \otimes q^{\gamma} + C_{G}
      \otimes G^{\,\gamma})                       \nonumber  \\
   & & \mbox{} \hspace*{1.3cm} +  \frac{\alpha}{2\pi} e_{q}^{2} C_{\, \gamma}
      \Big\} + \frac{1}{x} F_{2,\, h}^{\,\gamma} \:\: ,
\ea
where the summation extends over the light $u$, $d$ and $s$ quarks. The heavy
flavour contribution $ F_{2,\, h}^{\,\gamma} $ has recently been calculated to
second order in $ \alpha_{S} $ \cite{LRSN}, see also \cite{vN94}. $ C_{q,G} $
are the usual hadronic coefficient functions, and for the commonly used \MSb\
factorization one has $C_{\gamma} = 6 \, C_{G} $ \cite{BB}. The latter `direct'
photon term, however, causes difficulties in this scheme, since it leads to a
large LO/NLO difference for the pointlike contribution $ F_{2,\, PL}^{\,\gamma}
$ shown in Fig.\ 1. Especially, it is negative and divergent at large $x$:
\bq
  C_{\gamma ,\, \overline{MS}} \simeq 3\, \big[ \ln (1-x) -1 \big] \:\:
  \mbox{ for } \:\: x \rightarrow 1 \:\: .
\eq
This apparent perturbative instability in $ F_{2}^{\,\gamma} $ has to be
compensated by the NLO \MSb\ input parton densities, which therefore are forced
to be very different from the LO ones. In particular, a physically motivated
VMD input shape, vanishing for $ x \rightarrow 1 $, cannot be employed in NLO
here.

These problems are avoided by working in the \DISg\ factorization scheme
introduced in \cite{GRVt}. Here, the $C_{\gamma} $ term is (re-)absorbed in the
quark densities, i.e.\ the constant and the logarithmic parts of (1) are not,
as in (2), separated for the definition of $ q^{\,\gamma} $. The connection to
the usual \MSb\ scheme is
\bq
  q^{\,\gamma}_{DIS_{\gamma}} = q^{\,\gamma}_{\, \overline{MS}}
       + \frac{\alpha}{2\pi} e^{2}_{q} C_{\gamma ,\, \overline{MS}}
       \:\: , \:\:  C_{\,\gamma ,\, DIS_{\gamma}} = 0  \:\; ,
\eq
and the coefficient functions $ C_{q} $ and $ C_{G} $ as well as the definition
of the gluon density remain unchanged, in contrast to the hadronic DIS scheme
\cite{AEM}. Thus in \DISg\ $F_{2}^{\,\gamma} $ retains the usual hadronic \MSb\
form without a direct term also at NLO, resulting in a good LO/NLO stability
of $ F_{2,\, PL}^{\,\gamma} $ (Fig.\ 1). Hence in this scheme perturbatively
stable, physically motivated inputs for the photonic parton densities, e.g.
from\ VMD as in \cite{GRVph,GRVap}, can be used in NLO.

Due to the non-universality of the coefficient function $C_{\gamma} $,
$ F_{2}^{\,\gamma} $ has been assigned a special role in the redefinition (9)
of the quark distributions. The first (pragmatic) motivation for doing so is
that $ F_{2}^{\,\gamma} $ is the only photonic DIS structure function measured
so far and in the near future. The second (theoretical) reason stems from the
analogy with the hadronic DIS scheme, where $ F_{2} $ is special because of its
connection with the charge sum rules \cite{AEM}.

An additional advantage of the \DISg\ scheme, improving the perturbative
stability of the evolution, is that the strongest \MSb\ poles for $ x \ra 0 $
in the NLO photon-parton splitting functions $ k_{i}^{(1)} $ disappear:
\[
   k_{q}^{(1)} \!\sim  \Bigg\{ \!\! \begin{array}{r} \ln ^{2} x + \ldots \\
       2 \ln x + \ldots \end{array}  \! , \:
   k_{g}^{(1)} \!\sim  \Bigg\{ \!\! \begin{array}{rl} 1/x + \ldots \!\!\! &
        $ \MSb $ \!\!\! \\ -3 \ln x + \ldots \!\!\! & $ \DISg $ \!\!\!
        \end{array}
\]
\vspace{-0.2cm}\hfill (10) \\[0.05cm]
\addtocounter{equation}{1}
A corresponding cancellation takes place in the analogous timelike case, where
it is even more important: besides the corresponding structure functions, also
the photonic gluon fragmentation function $D_{g,\, PL}^{\,\gamma} $ is strongly
negative at small and medium $x$ in the \MSb\ scheme \cite{Aurfr}. These
problems are removed by using the timelike \DISg\ factorization \cite{GRVfr}.

An equivalent \MSb\ formulation of the above solution to the $ C_{\gamma}
$-problem is of course possible and has been approximately employed in
\cite{GS}. It can be written as a modification (`$\overline {\mbox{PL}}$') of
the pointlike part in (5) by an additional `technical' NLO quark input,
\bq
   q_{\,\overline{PL}}^{\gamma}(Q_{0}^{2}) = - \frac{\alpha}{2\pi} e^{2}_{q}
           C_{\gamma ,\, \overline{MS}} \:\: .
\eq
This leads to $ F_{2,\, \overline{PL}}^{\,\gamma}(Q_{0}^{2}) = 0 $ and thus
allows for similar `physical' inputs to $ \vec{q}_{had}^{\,\,\gamma} $ at
LO and NLO.
The resulting quark densities, however, depicted in Fig.\ 2 together with
their pointlike LO and \DISg\ counterparts, exhibit a rather unphysical shape.
%
\begin{figure}[tb]
\vspace{10cm}
\caption{The pointlike singlet quark density $ \Sigma_{PL}^{\gamma} $ in LO and
in NLO for the \DISg\ scheme compared to the physically equivalent \MSb\
distribution ($\overline {\mbox{PL}}$) with the input (11). The parameters
$ Q_{0}^{2} $ etc.\ are as in Fig.\ 1.}
\end{figure}

In a consistent NLO calculation, where e.g.\ the input (11) is evolved and
convoluted only with LO quantities, the above \MSb\ procedure is strictly
equivalent to the \DISg\ treatment for all physical quantities. On the other
hand, as soon as not all higher order $ {\cal O}(\alpha_{S}) $ terms are
discarded, the \MSb\ factorization turns out to be much less well-behaved than
the \DISg\ scheme. This is obvious from Fig.\ 3, where the results of a
`slightly' inconsistent calculation, in which only the additional $ {\cal O}
(\alpha_{S}) $ terms arising from the convolutions like $ \alpha_{S} C_{q}
\otimes q^{\gamma} $ in (7) have not been omitted, are compared to the
consistent result for both schemes. Consequently, in particular for
applications
 in the \MSb\ scheme, parametrizations of NLO photonic parton densities should,
as in \cite{GRVph}, supply also the leading part of $ \vec{q}^{\,\,\gamma}
_{NLO} $ in (6) ($ \ne \vec{q}^{\,\,\gamma}_{LO}$) necessary for consistent
calculations.
\begin{figure}[tb]
\vspace{10cm}
\caption{The consistent $ F_{2,\, PL}^{\,\gamma} $ of Fig.\ 1 in comparison
with inconsistent calculations, where spurious $ {\cal O}(\alpha_{S}) $ terms
from the convolutions in (7) have not been cancelled.}
\end{figure}

Very recently, an alternative approach has been suggested \cite{AFG}, where
a different (universal) technical input has been constructed by a detailed
analysis of the momentum integration of the Box-diagram:
\bq
   q_{\,PL^{\prime}}^{\gamma}(Q_{0}^{2}) = - \frac{\alpha}{2\pi} e^{2}_{q}
      C^{\prime}_{\gamma} \:\: , \:\:
   G_{\,PL^{\prime}}^{\gamma}(Q_{0}^{2}) = 0
\eq
with $ C^{\prime}_{\gamma} \simeq 3\,\ln (1-x) $ for $ x \rightarrow 1 $.
The resulting modified pointlike structure function $ F_{2,\, PL^{\prime}}
^{\,\gamma} $, also presented in Fig.\ 1, remains negative at large $x$ due to
the uncompensated -1 in (8) only very close to $ Q_{0}^{2} $.  However, a
significant dissimilarity to the LO result remains, which demands compensation
by sizeably different LO/NLO hadronic inputs.

Finally, it should be mentioned that the theoretical uncertainty of NLO photon
structure calculations has been investigated recently \cite{LRSN,AFG}. It has
been found that a variation of the renormalization and factorization scale
$ M^{2} $ from $ Q^{2}/4 $ to $ 4\, Q^{2} $ leads to an increase of $
F_{2} ^{\,\gamma} $ by less than 15\% (80\%) and 5\% (40\%) at $ Q^{2} = 6
\mbox{ GeV}^{2} $ and $ Q^{2} = 50 \mbox{ GeV}^{2} $ in NLO (LO), respectively
\cite{LRSN}. Thus theory is ready for significant comparisons with future
high-$ Q^{2} $ precise data, e.g.\ from LEP$\, 200 $.
\section{Photonic Parton Densities}
In order to specify the photon's parton distributions, one has to choose a
reference scale $ Q_{0}^{2} $ and to fix the perturbatively uncalculable
boundary conditions $ q^{\gamma}(Q_{0}^{2})$, $ G^{\,\gamma}(Q_{0}^{2}) $ for
the evolution equations (4). Experimentally, $ q^{\gamma} $ (mainly $
u^{\gamma} $) is known at $ 0.05 \stackrel{<}{{\scriptstyle \sim}} x
\stackrel{<}{{\scriptstyle \sim}} 0.8 $ from $ F_{2}^{\,\gamma} $ measurements
\cite{PLUTO,JADE,TASSO,TPC,CELLO,AMY,OPAL} with an accuracy of about $ 10
\div 20 \, \% $.
On the other hand, little is known about $ G^{\,\gamma} $ up to now:
there is evidence for $ G^{\,\gamma} \ne 0 $, and an extremely huge and hard
gluon (LAC$\, 3$, see below) has been ruled out by jet production data
\cite{TOPAZ,AMYj,H1j,ZEUSj}, see also \cite{Fo94}. The prospects on a
measurement of the gluon density are discussed in \cite{St94}. This lack of
experimental information is especially serious here since, as explained above,
there is no energy-momentum sum rule relating $ q^{\gamma} $ and $ G^{\gamma}
$.

VMD provides a connection between the quark and gluon inputs by the hadronic
momentum sum rule. Moreover, by SU(6) arguments, $ G^{\,\gamma} (Q_{0}^{2}) $
can then be estimated from the pionic gluon density $ G^{\,\pi}$, which has
been determined from direct photon production $ \pi p \ra \gamma X $
\cite{WA70,ABFKW}.
However, for the traditionally chosen input scales, $ Q_{0}^{2} \geq 1 \mbox{
GeV}^{2} $, a pure VMD ansatz is known to be insufficient. An additional hard
quark component has to be supplemented in order to achieve agreement with the
$ F_{2}^{\,\gamma} $ data at larger $ Q^{2} $ \cite{GGR,VS}.  In contrast to
the quarks, where the Born-Box (1) can provide a natural additional input, a
corresponding natural choice beyond VDM does not exist for the gluon density.

In view of this situation two approaches have been used. First, one can keep
$ Q_{0} \geq 1 \mbox{ GeV} $, fit the quark densities to $ F_{2}^{\,\gamma} $
data and `guess' the gluon input. This method has been adopted in \cite{DG}
and, more recently, in \cite{GS,LAC}. The second possibility is to try to
maintain the VMD idea and to start the perturbative evolution at a very low
scale $ Q_{0} < 1 \mbox{ GeV} $ \cite{GRVph,AFG,Aurph,ScSj}.
In the following, the recent parametrizations are discussed. All of them use
a QCD scale parameter of $ \Lambda = 200 \mbox{ MeV} $ for four active flavours
in LO and NLO. The PEP, PETRA and TRISTAN data on $ F_{2}^{\,\gamma} $
available when most of these parametrizations were done are shown in Fig.\ 4.
Since then, also first results from LEP have been published \cite{OPAL}, see
also \cite {Mi94}. The quark and gluon densities are compared in Fig.\ 5 and 6,
respectively.
\begin{figure*}[tp]
\vspace{10cm}
\caption{The 1991 world data on $ F_{2}^{\gamma} $ \protect\cite
{PLUTO,JADE,TASSO,TPC,CELLO,AMY} compared to the results obtained from the
LO and NLO GRV parametrizations, where the charm contribution has been
calculated via the lowest order Bethe-Heitler process \protect\cite{GRVph}.
The controversial \protect\cite{Field} low-$Q^{2}$ large-$x$ TPC/$2\gamma$
data points close to the resonance region (open circles) have not been used
for the determination of $\kappa $ in (13).}
\end{figure*}

\vspace{0.4cm}
\nd
{\em LAC (only LO)} \cite{LAC}:
\vspace{0.1cm}

The input scales are $ Q_{0}^{2} = 4 \mbox{ GeV}^{2} $ for sets 1, 2 and
$ Q_{0}^{2} = 1 \mbox{ GeV}^{2} $ for set 3. The boundary conditions have been
fitted to the data of Fig.\ 4 at $ Q^{2} > Q_{0}^{2} $ without any recourse to
physical constraints on the quark flavour decomposition and on the gluon
density. This has partially led to rather unphysical results even in the quark
sector, e.g.\ $ u^{\gamma} = 1.2 \alpha $, $ d^{\,\gamma} = 0.95 \alpha $,
$ s^{\gamma} = 2.8 \alpha $ and $ c^{\gamma} = 1.6 \alpha $ for LAC$\, 1$ at
$ x = 0.1 $ and $ Q_{0}^{2} = 10 \mbox{ GeV}^{2} $. Moreover, such a procedure
gives rise to wild reactions of the fitted gluon density on fluctuations and
offsets in the data:
in LAC$\, 1$ and LAC$\, 2$, $ G^{\,\gamma} $ is very large at $ x < 0.1 $ due
to the small-$x$ interplay of PLUTO data \cite{PLUTO} at low $ Q^{2} $ with
JADE \cite{JADE} and preliminary CELLO \cite{CELLO} results at larger $ Q^{2} $
in the fit. Likewise, the extremely huge and hard LAC$\, 3$ gluon is driven by
the inclusion of the large-$x$ small-$ Q^{2} $ TPC/2$\gamma $ data \cite{TPC},
which lie close to the resonance region and are therefore subject to
potentially large experimental correction factors and higher twist
contributions \cite{Field}. In any case the LAC gluons clearly illustrate that
present $ F_{2}^{\,\gamma} $ data do not provide useful constraints on
$ G^{\,\gamma} $.
\begin{figure*}[tp]
\vspace{9.5cm}
\caption{Photonic $u$-quark densities at LO and NLO. The LO distributions
\protect\cite{GRVph,GS,LAC} have been fitted to 1991 $ F_{2}^{\,\gamma} $ data,
for the different assumptions and data sets used see the discussions in the
text. The NLO results \protect\cite{GRVph,GS,AFG} are presented in the \MSb\
scheme. Here only the GRV distribution has been directly fitted to data.}
\end{figure*}
\begin{figure*}[htbp]
\vspace{10cm}
\caption{Photonic gluon distributions in LO and NLO. The spread of the LO
curves \protect\cite{GRVph,GS,LAC} exaggerates the present experimental
uncertainty on $ G^{\,\gamma} $, since LAC$\, $3 has been ruled out
\protect\cite{TOPAZ,AMYj}. On the other hand, the similarity of the NLO results
\protect\cite{GRVph,GS,AFG} is due to common VMD prejudices and not enforced by
data.}
\end{figure*}

\vspace{0.4cm}
\nd
{\em GRV (LO/NLO)} \cite{GRVph}:
\vspace{0.1cm}

The gluon and sea quark densities of the proton and the pion have been
generated from a valence-like structure at a common, very low resolution scale,
$ \mu \simeq 0.55 \, (0.5) \mbox{ GeV} $ in NLO (LO) \cite{GRVpr,GRVpi}.
First, this procedure has the advantage that few parameters in the gluon/sea
sector are needed. The pionic gluon density, being a parameter-free prediction
in the simplest version of this approach, shows very good agreement with the
results from direct photon production \cite{WA70,ABFKW}. Secondly, predictions
at small $x$ arise, where the distributions are determined by the QCD
dynamics and do virtually not depend on ambiguous input assumptions.  HERA
results on $ F_{2}^{\, p} $ \cite{ZEUSf,H1f} exhibit excellent agreement with
these predictions, especially when the charm contribution is properly taken
into account \cite{GRS}.

Therefore it is natural to utilize these low scales with $ \alpha_{S}/\pi
\approx 1/4 $ also in the photon case for a pure VMD ansatz in LO and NLO
(\DISg ):
\bq
  (q^{\gamma},G^{\gamma})(\mu^{2}) = \kappa \, (4\pi \alpha /
     f_{\rho}^{2})\, (q_{\pi^{0}},G_{\pi^{0}})(\mu^{2}) \:\: ,
\eq
with the coupling $ f_{\rho}^{2}/(4\pi) = 2.2 $ and $ 1 \stackrel{<}
{{\scriptstyle \sim}} \kappa \stackrel{<}{{\scriptstyle \sim}} 2 $ \cite{BeWa}.
In (13) the pion
distributions of \cite{GRVpi} have been used instead of the experimentally
unknown $ \rho $ densities. $\kappa $ accounts in a simple way for the higher
mass vector mesons $ \rho $, $ \omega $ and $ \phi $. Since this parameter is
not sufficiently specified by VMD, it has been fitted to the
$ F_{2}^{\gamma} $ data in Fig.\ 4. Good fits are obtained for the very
reasonable values $ \kappa = 1.6 \: (2.0) $ in NLO (LO). Data does not favour
neither an additional hard input at $ \mu^{2} $ nor a later onset of the
pointlike contribution.  Thus a pure VMD input at $ Q_{0}^{2} = \mu^{2} $ is
in fact sucessful and allows for an approximate prediction of $G^{\gamma} $ in
this
`dynamical' approach. Uncertainties stem from the $\rho \ra \pi $ substitution
and from the pion-valence normalization \cite{ABFKW,SMRS} entering the
determination of $ \kappa $. While the former is difficult to quantify, the
latter amounts to about $ 20\, \% $.

It is interesting to note that the same scale $ Q_{0} = 0.5 \mbox{ GeV} $ for
a LO VMD input has been obtained from quite different $\gamma p $ total cross
section considerations \cite{ScSj}. Consequently, the resulting distributions
resemble the ones of \cite{GRVph}. In order to remain flexible on the VMD part,
the LO pointlike solution (5) has been parametrized seperately in \cite{ScSj}.

\vspace{0.4cm}
\nd
{\em GS (LO/NLO)} \cite{GS}:
\vspace{0.1cm}

The input scale is $ Q_{0}^{2} = 5.3 \mbox{ GeV}^{2} $, the average of the
low-$ Q^{2} $ PLUTO data set \cite{PLUTO}. The $ F_{2}^{\,\gamma} $
measurements at $ Q^{2} > Q_{0}^{2} $ in Fig.\ 4 have partially been used to
fit the free parameters indicated below. In view of the gluon ambiguity
discussed above, two sets are provided with different assumptions on
$ G^{\, \gamma} $. At LO, the inputs of sets 1 and 2 are parametrized as
\ba
   q^{\,\gamma}_{1,2} & = & \mbox{VMD} + \mbox{QPM}(m_{q})    \\
   G^{\,\gamma}_{1}   & = & \mbox{VMD}
         + \frac{2}{\beta_{0}} P_{gq}^{\, (0)} \otimes \Sigma^{\,\gamma}
   \: , \: G^{\,\gamma}_{2} =  \mbox{VMD} \:\: .                   \nn
\ea
The VMD part is treated as in (13), with the old pion distributions
from \cite{OR} being used. The effective quark masses $ m_{\, u,d,s} $ in the
parton model (QPM) expression (1) and $ \kappa $ in (13) are determined from
data.
The difference between $ G^{\,\gamma}_{1} $ and $ G^{\,\gamma}_{2} $ presented
in Fig.\ 5 is about twice as big as the LO pointlike contribution evolved
from the start scale $ Q_{0} = 0.5 \mbox{ GeV} $ employed in \cite{GRVph,ScSj}.
Corresponding NLO (\MSb ) densities have been constructed without a new fit by
enforcing the same $ F_{2}^{\,\gamma}(Q_{0}^{2}) $ as in LO together with
assumptions on the flavour decomposition.

\vspace{0.4cm}
\nd
{\em AFG (only NLO)} \cite{AFG}:
\vspace{0.1cm}

These distributions supersede the NLO ones of \cite{Aurph}, where a pure VMD
input was employed at LO and NLO (\MSb ) at $ Q_{0} = 0.5 \mbox{ GeV} $.
Here, the technical input (12) is supplemented by a VMD ansatz at $ Q_{0} =
0.7 \mbox{ GeV} \approx m_{\rho} $. No fit to data is performed.
Somewhat different from (13), a coherent sum of $\rho $, $\omega $ and
$ \phi $, is used, where identical valence and gluon
distribution in all three vector mesons and the pion have been assumed:
\bq
   u^{\gamma} \! = K \alpha \bigg[ \frac{4}{9} u^{\pi}_{val} + \frac{2}{3}
           u^{\pi} _{sea} \bigg] \, , \ldots ,\: G^{\,\gamma} \! = K \alpha
		\frac{2}{3} G^{\,\pi}
\eq
The pion densities are adopted from \cite{ABFKW} with the normalization $ K $
left free in the parametrization. The standard choice reads $ K = 1 $, with a
$ 50 \,\% $ variation allowed by high-$ Q^{2} $ data on $F_{2}^{\,\gamma} $.
\section{$\alpha_{S} $ and the photon structure}
The sensitivity of $F_{2}^{\,\gamma} $ to the QCD scale parameter $ \Lambda $
considerably decreases with increasing scale $ Q^{2}_{0} $, at which the
evolution (4) of the photonic partons is startet \cite{VS,FKP,Fra}. For $
Q^{2}_{0} \simeq 5 \mbox{ GeV}^{2} $, as in \cite{GS}, only a marginal effect
is left.
It interesting, however, to reconsider this issue employing
the `dynamical' approach, where the parton densities of
nucleon, pion and photon are related by valence-like inputs and VMD at a
common, very low resolution scale, $ \mu \simeq 0.55 \mbox{ GeV} $ for
$ \Lambda_{\overline{MS}}^{(4)} = 0.2 \mbox{ GeV} $ \cite{GRVph,GRVpr,GRVpi}.
Therefore, this procedure, carried out only for this fixed $ \Lambda $ so far,
has been repeated for different values of $\Lambda_{\overline{MS}} $ with
$ \mu (\Lambda ) $ fixed by proton structure information.
The free parameter in the photon sector, $ \kappa $ in (13), has been
fitted to the published results on $F_{2}^{\,\gamma} $ with a cut of
$ Q^{2} > 4 \mbox{ GeV}^{2}$ in order to suppress possible higher-twist
contributions. These fits exhibit a rather good
$\Lambda $-sensitivity in this dynamical scenario,
\bq
   \Lambda_{\overline{MS}}^{(4)} = (210 \pm 50_{exp.}) \mbox{ MeV}
   \:\: ,
\eq
where the $1\,\sigma $ error given in (16) arises from the total
experimental uncertainties.
A more detailed analysis including theoretical uncertainties will be presented
in a seperate paper \cite{AVpr}. Since more precise data are expected,
especially from LEP$\, 200$, prospects do not seem to be too dim for a
competitive $\alpha_{S} $ determination from $F_{2}^{\,\gamma} $.
%
%
%

\end{document}